\documentclass[12pt]{article}
\usepackage[utf8]{inputenc}
\usepackage[margin=1in]{geometry}
\usepackage[english]{babel}
\usepackage[natbibapa]{apacite}
\usepackage[title]{appendix}
\usepackage{amsmath,amsthm,indentfirst,physics,float,footnote,tablefootnote,graphicx,booktabs,enumerate,xcolor,setspace}
\newtheorem{theorem}{Theorem}
\newtheorem{lemma}[theorem]{Lemma}

\title{\textbf{Set-Asides in USDA Food Procurement Auctions}}

\author{Ni Yan \and WenTing Tao}

\date{Stanford University}

\begin{document}
\maketitle
\doublespacing
\begin{abstract}
We study the partial and full set-asides and their implication for changes in bidding behavior in first-price sealed-bid auctions in the context of United States Department of Agriculture (USDA) food procurement auctions. Using five years of bid data on different beef products, we implement weighted least squares regression models to show that partial set-aside predicts decreases in both offer prices and winning prices among large and small business bidders. Full set-aside predicts a small increase in offer prices and winning prices among small businesses. With these predictions, we infer that net profit of small businesses is unlikely to increase when set-asides are present.
\end{abstract}

\section{Introduction}

Promoting U.S. small businesses has long been an important objective of the U.S. government. Small Business Act of 1953 mandates that each federal agency should create ``an annual goal that presents, for that agency, the maximum practicable opportunity for small business concerns''.\footnote{15 U. S. C. §644(g)(1)(B)} Through programs and rules that prioritize small businesses over large ones, the U.S. government hopes to promote the overall growth and competitiveness of various types of small businesses. This paper focuses on the impact of partial and full set-asides policies favoring small businesses in the U.S. Department of Agriculture (USDA) food procurement auctions.

Every year, the USDA purchases billions of pounds of agricultural products from vendors across the country through first-price sealed-bid online multiunit auctions. In each auction, the items being procured appear as individual contracts that specify the product type, product quantity, delivery destination and delivery window. Winners are responsible for completing contracts (referred to as ``items'' hereinafter) to receive payment. Auctions happen around once a week for each food category such as beef, milk, fruits, etc. Partial or full set-asides are applied to product type, and guarantee that a certain percentage of total quantity procured for that product is awarded to small businesses. This percentage is subject to change based on USDA's assessment of small businesses' production capacity.

This paper focuses on beef procurement auctions in the years 2015 to 2019. Beef category is selected because we are able to identify a set of relatively similar beef products that each fall into different set-aside percentages across time and auctions. We include all the bids data available from the USDA's website, taking out years 2020 and 2021 to eliminate the potential confounding effect of the Covid-19 pandemic. We quantify the association between set-asides and the number of participants, offer prices and winning prices using weighted least squares regressions with set-aside percentage, bidder type, their interaction and other observables as covariates. Our findings motivate the question of whether small businesses benefit from partial set-aside policies used by USDA, as the existence of partial set-asides is associated with a decrease in both offer prices and winning prices among large and small businesses. Lower winning prices imply a smaller price margin for small businesses, while the data suggests it's unlikely that their probability of winning can increase enough to compensate for price drop and drive up their net profit. In full set-aside auctions, small businesses raise their prices as expected, but the increase is small because there are few large business competitors to begin with. 

The paper is organized as follows. Section II reviews existing literature on the topic of set-asides in procurement auctions. Section III describes in detail the format of the USDA procurement auctions and how set-asides are implemented. Section IV provides a simplified behavioral model for auctions of homogeneous items with partial set-asides. Section V conducts weighted least squares regressions of the number of bidders, offer prices and winning prices using USDA beef auction data. Section VI discusses potential issues and directions for future research. Section VII concludes.

\section{Literature Review}
Although set-asides policy is a popular topic in auction theory, most work has been focusing on evaluating changes in revenue and social surplus resulting from full set-asides, where large businesses are excluded from the auction. Some other work compares the efficiency of set-asides and subsidies in the form of discounted payments to small businesses. Opinions on the effectiveness of set-asides are mixed and depend on the setting. Little work has been done studying partial set-asides that allow large businesses into the auction but put a constraint on their winnings. This paper aims to fill the gap of partial set-asides policies in procurement auctions through empirical analysis.

\cite{athey2011} compare the effect of set-asides and subsidies on revenue and social surplus in U.S. Forest Service timber auctions. The timber auction adopts either open competition between small and large bidders, or full set-asides for small bidders. They estimate a parametric model to recover value distribution from observed bids in auctions without set-asides, and build a probabilistic model of bidder entry subject to entry cost and expected profit. Counterfactual analysis shows that switching from set-asides to open competition would introduce 1.7 large bidders and lose 1.9 small bidders, with the fraction of sales won by small bidders dropping. Auction revenue and bidder surplus increase by 6\% and 21\% respectively. As summarized in this paper, understanding the impact of set-asides on participation of potential bidders is a crucial step in the analysis of subsequent changes in bidding strategy.

\cite{nakabayashi2013} studies full set-aside auctions in the context of Japanese public construction projects focusing on the impact on government budgets, taking into consideration changes in bidder participation. This paper concludes that the set-aside programs increase small businesses' participation in procurement auctions by around 40\%, which is shown to be high enough to reduce government procurement costs.

The two papers above are both based on an environment with flexible entry, yet the results are in different directions. It seems important to understand features of the underlying industry before thinking about the effect of set-asides. \cite{macdonald2002} study the effect of an additional bidder on competitive pricing in the USDA procurement auctions of flour, pasta and oil in 1992 to 1996. Regressing winning prices on ingredient price, product, quantity and competition, they find that an additional bidder is associated with a much larger price drop when there are fewer competitors. We incorporate this nonlinear effect of the number of bidders in our model and find similar results. They also compare USDA realized prices with private sector wholesale prices for comparable products and services, and find that USDA prices are generally lower by 20-40\%. The large price gap is a result of the USDA aiming to find the lowest short-run marginal cost among vendors. This might account for our observation of few large bidders but an abundance of small bidders in our data. USDA auctions could be unattractive to large bidders because they operate on a large scale and use more of long-term contracts. Inspired by their analysis we include monthly wholesale beef price as a proxy for the option to sell outside of the USDA auctions.

\section{Description of USDA Food Procurement Auction and Set-Asides}

The USDA spends around two billion dollars every year on two billion pounds of domestically produced and processed food, including grain, meat, vegetable, fruit, etc., to be delivered to schools, food banks, and households. About 400 million dollars is spent in the U.S. on beef products annually, mostly on frozen ground beef and patties. The USDA's Agricultural Marketing Service (AMS) operates first-price sealed-bid multiunit auctions through their online system WBSCM\footnote{Web-Based Supply Chain Management.}. Only qualified vendors registered in their system can participate in their online auctions. USDA conducts a plant survey to evaluate the technical, production, and transportation capabilities of applicants following its own product specification and technical requirement that dictates food composition, package size, and package material. This paper focuses on nine ground beef products which we will discuss in more detail in Section V, and for these products only 17 vendors ever occurred in a five-year span, suggesting that bidder entry is quite limited and infrequent. Figure 1 shows that the number of active bidders across time is quite stable.\footnote{We identify the earliest and latest auction each vendor participated and count the time in between as their active period. The surge at the beginning in Figure 1 is simply due to lack of earlier data.}
\begin{figure}[H]
    \centering
    \includegraphics[width=0.8\textwidth]{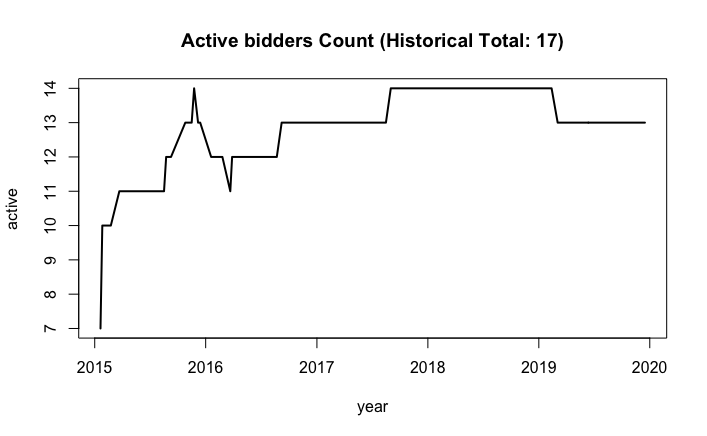}
    \caption{Number of potential bidders}
\end{figure}

Procurement announcements are first published in AMS newsrooms, followed by a detailed solicitation letter posted on USDA's website. Each solicitation consists of a list of items that specifies a USDA product, the quantity in pounds, delivery destination (city, state and zip code), and delivery window. Qualified vendors submit an independent price per pound for each item, and any maximum quantity constraint they have for certain product and delivery window.\footnote{For example, if a vendor only has two trucks of ground beef and is indifferent among all delivery windows, the vendor can bid for all delivery windows and report a capacity constraint on at most two trucks for all windows.} To avoid confusion we use the word product to refer to the product group that an item belongs to hereinafter.\footnote{A solicitation can involve multiple products, and each product group can involve multiple items. Though it is called product group, it is actually a very specific type of product defined by USDA, e.g. BEEF 100\% PTY 85/15 FRZ 2.0MMA CTN-40 LB, which means 100\% frozen beef patty with fat content 85/15, packed in a 40-pound carton and meeting the 2.0 credit level of meat/meat alternate.} In beef procurement auctions, which run weekly, there is usually one or two weeks for bidders to submit their bids before the auction closes. The delivery window is usually a two-week span, and winning bidders are responsible for delivering products to specified central warehouses before they receive payment.

Each solicitation may contain one or multiple product types with different quantities. Following FAR 52.219-6 Notice of Small Business Set-Aside, the USDA implements a percentage set-aside policy, given that in the Animal (except Poultry) Slaughtering industry (NAICS Code 311611) 1,000 employees or less is the standard of small business. Set-aside percentage is announced for the solicitation and is applied at the product type level, meaning that all items of that product type are considered when calculating the percentage quantity that is awarded to small businesses. For instance, consider a solicitation where all items are 40,000 lbs. If there are five boneless fresh beef items, four frozen ground beef items, and three beef patty items, with set-aside percentages 0\%, 50\% and 100\% respectively, then at least two frozen ground beef items and all three beef patty items should be awarded to small business bidders if feasible, and there's no restriction on boneless fresh beef items. In particular, applying the set-aside percentage at a product type level guarantees that such a solicitation with three product types is theoretically equivalent to three independent solicitations of one product type each. AMS also has some discretion to cancel orders that end up with unreasonably high prices. Because AMS's price ceiling is not public information, and cancellation due to high prices is rare in the data, these online auctions are effectively first-price sealed-bid multiunit auctions subject to linear constraints.

\section{A Simplified Behavioral Model}
To understand how partial set-asides affect the competition within and between large and small bidders, let's consider a simplified model that auctions one product type with identical items. Suppose there are $M$ items, $N_1$ small bidders and $N_2$ large bidders. A set-aside percentage of $\alpha=\frac{M_1}{M}$ dictates that at least $M_1$ items go to small bidders if feasible. Because items are identical, we focus on the pure strategy that vendors bid one price for all items. Suppose the best response functions of small and large bidders are $b_1(v_1)$ and $b_2(v_2)$ respectively where $v_1 \overset{iid}{\sim} F_1$ and $v_2 \overset{iid}{\sim} F_2$. Assume that $b_1(\cdot)$ and $b_2(\cdot)$ are increasing and invertible, and their inverse are $c_1(\cdot)$ and $c_2(\cdot)$ respectively.

The simplest case is a $\alpha$ set-aside auction, i.e. $M_1 = \alpha M$, with $N_1=2$ small bidders, $N_2=1$ large bidder and no capacity constraint or price ceiling.\footnote{More than one small bidder guarantees that bidding at price ceiling is a dominated strategy.} If the large bidder has the lowest price, then she gets any $(1-\alpha)M$ items and the cheaper small bidder gets the remaining $\alpha M$ item. If the large bidder doesn't have the lowest price, then only the cheapest small bidder gets all $M$ items.

We compute the expected profit of a small bidder with value $v_1$ and price p as the following:
\begin{align*}
    \pi_1(p,v_1)&=(p-v_1)\cdot P(p < p_{other\,small}) \cdot [\alpha M\cdot P(p>p_{large})+M\cdot P(p\leq p_{large})] \\
    &=M\cdot(p-v_1)\cdot P(p < p_{other\,small}) \cdot [\alpha \cdot P(p>p_{large})+ (1-P(p>p_{large}))] \\
    &=M\cdot(p-v_1)\cdot P(p < b_1(v_{other\,small})) \cdot [1- (1-\alpha)\cdot P(p>b_2(v_{large}))] \\
    &=M\cdot(p-v_1)\cdot P(c_1(p) < v_{other\,small}) \cdot [1- (1-\alpha)\cdot P(c_2(p) > v_{large})] \\
    &=M\cdot(p-v_1)\cdot [1-F_1(c_1(p))] \cdot [1-(1-\alpha)F_2(c_2(p))]
\end{align*}

And the expected profit of a large bidder with value $v_2$ and price p is the following:
\begin{align*}
    \pi_2(p,v_2)&=(1-\alpha)\cdot M\cdot(p-v_2)\cdot P(p < both\; p_{small}) \\
    &=(1-\alpha)\cdot M\cdot(p-v_2)\cdot P(p < b_1(v_{small}))^2 \\
    &=(1-\alpha)\cdot M\cdot(p-v_2)\cdot P(c_1(p) < v_{small})^2 \\
    &=(1-\alpha)\cdot M\cdot(p-v_2)\cdot [1-F_1(c_1(p))]^2
\end{align*}

Taking first order condition of a small bidder wrt. $p$ gives us:
\begin{align*}
     &0 = [1-F_1(c_1(p)] \cdot [1-(1-\alpha)F_2(c_2(p))] + \\
     &\quad\quad(p-v_1)\cdot \{[-F_1(c_1(p))\cdot f_1(c_1(p))\cdot c_1'(p)] \cdot [1-(1-\alpha)\cdot F_2(c_2(p))] \\
     &\quad\quad\quad\quad\quad\quad+[1-F_1(c_1(p))] \cdot (1-\alpha)[-F_2(c_2(p))\cdot f_2(c_2(p))\cdot c_2'(p)]\} \tag{1}
\end{align*}

In the first order condition above, for any $v_1$, $p=b_1(v_1)$ should solve the equation. Equivalently, for any $p$, $v_1=b_1^{-1}(p) = c_1(p)$ also solves the equation (1).

Taking first order condition of a large bidder wrt. $p$ gives us:
\begin{align*}
     0 &= [1-F_1(c_1(p))]^2 + (p-v_2)\cdot 2\cdot [1-F_1(c_1(p))] \cdot [-f_1(c_1(p))\cdot c_1'(p)] \\
     0 &= 1-F_1(c_1(p)) + (p-v_2)\cdot 2\cdot [-f_1(c_1(p))\cdot c_1'(p)]
\end{align*}
where the term $1-F_1(c_1(p))$ can be factored out and dropped because $1-F_1(c_1(p))=0$ would imply zero profit for the large bidder. With the same reasoning as in the first order condition of small bidders, equation (2) should hold for all p if we replace $v_2$ by $c_2(p)$.

Therefore, the solutions $c_1(\cdot)$ and $c_2(\cdot)$ are characterized by the following first-order differential equation system:
\begin{align*}
    c_1(p) &= p-\frac{[1-F_1(c_1(p)] \cdot [1-(1-\alpha)F_2(c_2(p))]}{\{[F_1(c_1(p)\cdot f_1(c_1(p))\cdot c_1'(p)] \cdot [1-(1-\alpha)F_2(c_2(p))]+} \\
     &\quad\quad\quad\;\;[1-F_1(c_1(p))] \cdot (1-\alpha)[F_2(c_2(p))\cdot f_2(c_2(p))\cdot c_2'(p)]\} \tag{3} \\
    c_2(p) &= p- \frac{1-F_1(c_1(p))}{2f_1(c_1(p))\cdot c_1'(p)} \tag{4}
\end{align*}

Notice that $c_1(p)$ depends on $\alpha$ directly and $c_2(p)$ only depends $\alpha$ through $c_1(p)$. $v_2=c_2(p)$ can be expressed with $c_1(p)$ and it's derivative, and $\frac{1-F_1(c_1(p))}{2f_1(c_1(p))\cdot c_1'(p)}$ is the markup of large bidders. To do a thought experiment, suppose $F_1$, the distribution of small bidders' value, is uniform, then $f_1(c_1(p))$ is a constant. If a change in $\alpha$ would result in an increase in both $c_1(p)$ and $c_1'(p)$ for every p, then the markup at every p will decrease since the numerator falls, the denominator rises, and they are both positive. It then follows that $c_2(p)$ also increases for every p. Inverting $c_1(\cdot)$ and $c_2(\cdot)$ gives us the conclusion that under such assumptions it's possible that $b_1(v_1)$ and $b_2(v_2)$ move in the same direction. In the next section we'll use weighted least squares models on USDA bids data to predict changes in bidders' prices empirically.

\section{Results}
The USDA maintains a list of qualified vendors with their large or small business statuses, and occasionally adjusts set-aside percentage for certain product types. We obtained data of bid prices and awards of individual auctions under the beef category on the USDA website from Jan 01, 2015 to Dec 31, 2019. There are 23 types of products auctioned in this period. We first discard products that don't involve a change in set-aside percentage which leaves us with seven types of products. Because the exact same products packaged in different sizes will be classified as different products by the USDA, we add back two such products to enlarge the dataset, and control for package size in regressions.\footnote{For example, BEEF FINE GROUND LFT OPT FRZ PKG-40/1 LB and BEEF FINE GROUND LFT OPT FRZ CTN-40 LB only differ in package size. The former is packed in forty 1-pound packages and the latter in four 10-pound packages.} The data contains 128 auctions on 9 products, corresponding to 13,192 items and 96,474 bids. Three set-aside percentages are involved: 0\%, 50\% and 100\%, associated with 762, 12215 and 215 items respectively.\footnote{The 9 products are (1)BEEF 100\% PTY 85/15 FRZ 2.0MMA CTN-40 LB (2)BEEF 100\% PTY 90/10 FRZ 2.0MMA CTN-40 LB (3) BEEF PATTY LEAN FRZ CTN-40 LB (4) BEEF SPP PTY 85/15 FRZ 2.0 MMA CTN-40 LB (5) BEEF COARSE GROUND FRZ CTN-60 LB (6) BEEF FINE GROUND FRZ CTN-40 LB (7) BEEF FINE GROUND FRZ PKG-40/1 LB (8) BEEF FINE GROUND LFT OPT FRZ CTN-40 LB (9) BEEF FINE GROUND LFT OPT FRZ PKG-40/1 LB}

Table 1 shows summary statistics of auctions and items grouped by set-aside percentage. Small bidder pool and large bidder pool in Table 1 refer to the number of small or large bidders that ever occurred in that set-aside group. There are some differences in the size of bidder pool and the average number of bidders because set-asides apply to product type level. Therefore we control for the number of bidders and its quadratic term in all regressions. Fifty percent set-aside is the most common in the data because it is related to a product that the USDA procures frequently. We control for a variety of product and item characteristics such as demand level, quantity, the number of bidders, product type, product value, etc., to eliminate bias resulting from differences among products because the set-aside percentage is not randomly assigned to the products.

\begin{table}
    \centering
    \begin{tabular}{r|ccc}
    \toprule
    & SA=0\% & SA=50\% & SA=100\% \\\midrule
    number of auctions & 43 & 127 & 39 \\
    number of items & 762 & 12215 & 215 \\
    number of bids & 1659 & 94218 & 597 \\
    Small bidder pool & 4 & 14 & 8 \\
    Large bidder pool & 1 & 5 & 0 \\
    Average number of small bidders & 1.835 & 6.656 & 3.147 \\
    & (0.466) & (1.602) & (1.042) \\
    Average number of large bidders & 0.507 & 1.788 & 0 \\
    & (0.533) & (1.082) & (0) \\
    Mean Offer price & 2.966 & 2.587 & 2.708 \\
    & (0.593) & (0.354) & (0.374) \\
    Mean Winning price & 2.742 & 2.469 & 2.577 \\
    & (0.468) & (0.332) & (0.322) \\
    Mean Item Quantity & 32841 & 40707 & 38022 \\
    & (9088) & (4412) & (5086) \\
    Exist in all years & Yes & Yes & Yes \\
    \bottomrule
    \end{tabular}
    \caption{Summary Statistics}
\end{table}

\subsection{Number of Bidders}
Figure 2 left panel draws a boxplot of the number of bids from large bidders and from small bidders for each item and is grouped by set-aside percentage. Figure 2 middle panel draws a jittered scatter plot of the share of quantity won by large and small bidders in each auction and product type, also grouped by set-aside percentage. Figure 2 right panel visualizes the percentage of total quantities won by large and small bidders grouped by set-asides. First, we look at products with no set-asides and see very few large and small bidders. Their winning shares in each auction and product type are both spread out, but small bidders win much more often and get 95\% of total quantities in general. Then for products with 50\% set-asides, there are more large bidders and a lot more small bidders. Their winning shares are pushed to different sides of the 50\% set-asides bar, with occasional exceptions due to lack of small bidders' offers. Small bidders continue to dominate the auction, winning 87\% of total quantities. Lastly, we see a few small bidders for products with 100\% set-asides, and they win all quantities unless they can't supply that much. In general, we see that small bidders win a large share of quantities in any set-asides. It implies that even if it is the case that set-asides increase the probability of a small bidder winning, such increase on average cannot be large because small bidders are already doing well. In Section VI we will link this idea with our findings on bidders' price changes in Section 5.2 and 5.3 to comment on small businesses' potential revenue change.
\begin{figure}
    \centering
    \includegraphics[width=\textwidth]{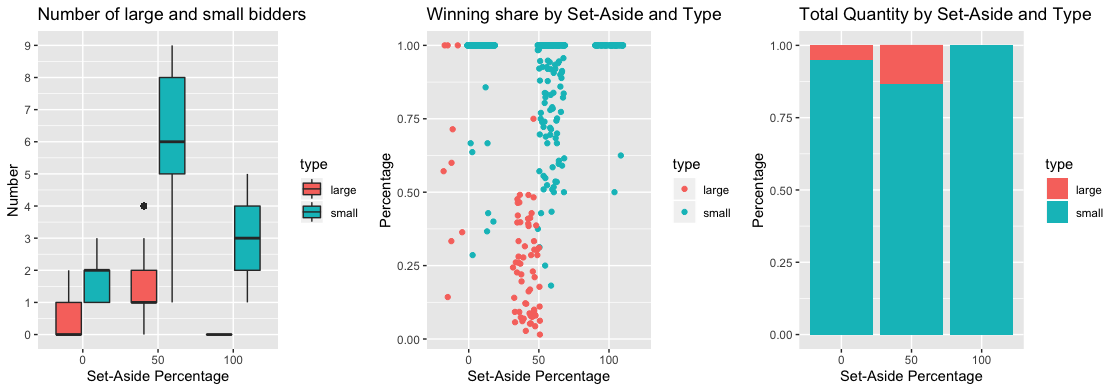}
    \caption{Number of bidders and Share of Winning}
\end{figure}

To quantify the effect of set-asides, we estimate a weighted least squares model of the number of small or large bidders for each item on set-aside percentage, bidder type, their interaction, and other covariates (equation 5). For each item we count the number of large bidders and the number of small bidders to generate two observations in the response variable. We include bidder type as a covariate to indicate which type of bidder is being counted in the response variable. Because the effect of set-aside is expected to be nonlinear, set-aside percentage is considered as a categorical variable in all regression models. We also control for demand level, which is computed as the total quantity of the same product in million pounds requested in that solicitation. Demand is taken into consideration because large orders may attract more bidders who want a slice of the cake, and a quadratic term of demand is included to capture potential non-linearity. Product and year\footnote{Year is treated as a categorical variable in all regression models so that we don't assume a linear time trend.} fixed effects are controlled for to capture the intrinsic differences in competition level for different products, as well as fluctuation in bidders' activity. Lastly, an indicator variable SDVOSB considers whether a 10\% service disabled veteran owned small business set-aside applies to the product.\footnote{Unfortunately we do not have data on service disabled veteran owned small businesses status. A web search suggests that two vendors might be veteran-owned today, but historical records are unavailable. Because 10\% is relatively small, we decide to include an indicator variable of SDVOSB in all regression models. SDVOSB is not additive to the set-asides we are concerned about, in the sense that SDVOSB bidders count toward both set-aside quota.}
\begin{align*}
\text{Num of bidders}\sim &\text{ Set-Aside} + \text{Bidder Type} + \text{Set-Aside $\times$ Bidder Type} \\
&+\text{Demand} + \text{Demand}^2 +\text{Product}+\text{year}+ \text{SDVOSB}\tag{5}
\end{align*}

For each observation in the sample, we use the quantity of that item as its weight in the model. The quantity of most items is the weight of a standard truckload, usually 42,000 lbs, 40,000 lbs or 38,000 lbs depending on the product. However, some items in the data appears as less than a unit truckload for administrative reasons\footnote{For example, an auction may request half a truck of ground beef to be delivered to one place and another half to a different place in the same city. They would appear as two half-truck items in the data, but in practice bidders are required to treat them as a bundle and submit only one price. Using item quantity as weight perfectly ``merges'' them into one standard truckload item.}, so we use quantity as weights to equally represent every truckload. Table 2 reports the results using heteroskedasticity-consistent robust standard errors.\footnote{Coefficients on commodity fixed effect is left out.} The first two columns are results from weighted least squares and the last two columns are those from ordinary least squares without weighting. OLS coefficients are close to WLS coefficients, and we'll only discuss the latter in detail. The baseline of the interaction terms is the number of large bidders in no set-asides setting. On average there are 1.4 more small bidders than large bidders for items without set-asides. Items with a 50\% set-aside are associated with a 1.3 decrease in the number of large bidders and a 1.9 increase in the number of small bidders. Recall from Figure 2 that the median number of large and small bidders in this category is 1 and 6. Items with a 100\% set-aside are associated with a 0.11 increase in the number of small bidders. The smaller magnitude of absolute increase compared to 50\% set-asides is likely because the latter is more competitive to begin with. All the coefficients are significant and the signs are intuitive. Implementing set-asides discourages large bidders from participating likely because there's a lower chance of winning the auction and the set-asides attract small bidders who are advantaged. Demand ranges from 0.038 to 13.272 million pounds. Within this range, demand predicts an increase in the number of bidders of any type ranging from 0 to 1.1.

\begin{table}[H]
    \centering
    \begin{tabular}{r|cc|cc}
    \toprule
    & \multicolumn{2}{c|}{WLS} & \multicolumn{2}{c}{OLS} \\
    & Coefficient & Robust Error & Coefficient & Robust Error \\\midrule
    Constant & 0.6060 & (0.0353) & 0.6664 & (0.0344) \\
    Small & 1.4050 & (0.0272) & 1.3373 & (0.0255) \\
    SA50\%, Large & -1.3251 & (0.0879) & -1.3592 & (0.0852) \\
    SA50\%, Small & 1.8804 & (0.0865) & 1.9113 & (0.0846) \\
    SA100\%, Small & 0.1142 & (0.0395) & 0.1582 & (0.0366) \\
    Demand & 0.1506 & (0.0133) & 0.1479 & (0.0127) \\
    Demand$^2$ & -0.0050 & (0.0010) & -0.0046 & (0.0010) \\
    SDVOSB & 0.4747 & (0.0696) & 0.4843 & (0.0689) \\ \midrule
    $n=26169$ &\multicolumn{2}{c|}{$R^2=0.8134$} & \multicolumn{2}{c}{$R^2=0.8155$} \\ \bottomrule
    \end{tabular}
    \caption{Number of Large or Small Bidders}
\end{table}

\subsection{Offer Price}
Next we study the association between set-asides and offer prices submitted by large and small bidders. We use a weighted least squares model of the log of offer price regressed on set-aside percentage, bidder type, their interaction and other covariates (equation 6). We include demand for the product and total number of bidders for each item as covariates because prices should be affected by demand and supply relationship. Although bidders do not know the actual number of competitors when they submit their bids, they should have a good estimate of it because auctions are highly repetitive and the pool of qualified vendors is quite stable. Quadratic terms are used to capture non-linearity.

Because the data spans five years and the market price of beef products fluctuates, we introduce two other price data published by the USDA. One is monthly data of the wholesale value of beef. Because the USDA auctions are comparable to the wholesale process which also sells at large quantities to retailers, we match the log of beef wholesale value with auction items by year and month, and use it as a proxy for the opportunity cost to sell outside of the USDA auctions. The other information is the 2018-2019 annual market price of USDA products. As mentioned in the data section, we will consider USDA products that only differ in package size as the same product. Considering that package size affects cost, we use a product fixed effect, the log of 2019 USDA price, and a package indicator altogether to capture intrinsic price differences among products. Last, we use vendor fixed effect to control for the efficiency differences among bidders, and a SDVOSB indicator of service-disabled veteran owned small business set-aside. 
\begin{align*}
\text{log(Offer Price)}\sim & \text{ SA} + \text{Bidder Type} + \text{SA $\times$ Bidder Type} + \text{Demand} + \text{Demand}^2 \\
& + \text{nBid} + \text{nBid}^2  +\text{log(Monthly Wholesale Beef Price)} + \text{Product} \\
&+ \text{log(2019 USDA price)} + \text{Vendor}+ \text{Package}+\text{SDVOSB}\tag{6}
\end{align*}

Table 3 reports the results with heteroskedasticity-consistent robust standard errors. OLS coefficients are close to WLS coefficients which we'll discuss in detail. Taking the exponential of the coefficients gives the corresponding percentage change in offer price. A positive estimate of small bidder indicator suggests that for items without set-asides, the average offer price from small bidders is 6.5\% higher than those from large bidders. Interestingly, for items with 50\% set-aside, the offer prices of large bidders are expected to drop by 13.6\% on average, suggesting that large bidders bid more aggressively in reaction to their lower chance of winning. Similarly, 50\% set-aside predicts a larger drop of 21.7\% in the offer prices of small bidders on average. The decrease in offer prices among small bidders could be a response to large bidders' aggressive bidding. As we saw in the simplified behavioral model, in a 50\% set-aside setting only the cheaper small bidders actually benefit from set-asides. In addition, extra uncertainty is introduced to the allocation outcome in auctions with partial set-asides. Without set-asides, a small bidder who wants to sell her only truck of beef to a close destination only needs to be the lowest price for the targeted item. However, with a 50\% set-aside, this bidder wins if and only if the sum of her price and other bidders' low prices for all other items of the same product is the cheapest for the USDA. 

For items with a 100\% set-aside, offer prices of small bidders increase by 3.2\%, which is expected because there is no more competition from large bidders. The relatively small magnitude is also understandable because there are few large bidders compared to small bidders in USDA auctions. Total demand for a product predicts a roughly linear increase in offer prices that ranges from 0.06\% to 12\%. Number of bidders predicts a decrease in offer price as expected, because competition will drive down prices in first price auctions. A negative coefficient of the quadratic term implies that the impact of an additional bidder gets weaker when there are more bidders to begin with, which aligns with the results in \cite{macdonald2002}. The model predicts that adding one more bidder for a typical item is expected to bring down the offer price by about 3\%. All the coefficients discussed above are significant.
\begin{table}
    \centering
    \begin{tabular}{r|cc|cc}
    \toprule
    & \multicolumn{2}{c|}{WLS} & \multicolumn{2}{c}{OLS} \\
    & Coefficient & Robust Error & Coefficient & Robust Error \\\midrule
    Constant & -1.2210 & (0.0347) & -1.2399 & (0.0339) \\
    Small & 0.0630 & (0.0173) & 0.0566 & (0.0163) \\
    SA50\%, Large & -0.1457 & (0.0200) & -0.1625 & (0.0187) \\
    SA50\%, Small & -0.2443 & (0.0092) & -0.2533 & (0.0083) \\
    SA100\%, Small & 0.0318 & (0.0069) & 0.0354 & (0.0068) \\
    Demand & 0.0172 & (0.0005) & 0.0166 & (0.0005) \\
    Demand$^2$ & -0.0006 & (0.0000) & -0.0006 & (0.0000) \\
    Number of bidders & -0.0310 & (0.0013) & -0.0309 & (0.0013) \\
    Number of bidders$^2$ & 0.0013 & (0.0001) & 0.0013 & (0.0001) \\
    log(USDA price) & 1.6662 & (0.0301) & 1.7051 & (0.0299) \\
    log(wholesale price) & 0.7936 & (0.0048) & 0.7947 & (0.0047) \\
    SDVOSB & 0.0322 & (0.0016) & 0.0330 & (0.0016) \\ \midrule
    $n=96474$ &\multicolumn{2}{c|}{$R^2=0.3985$} & \multicolumn{2}{c}{$R^2=0.4058$} \\ \bottomrule
    \end{tabular}
    \caption{Log of Offer Price}
\end{table}

\subsection{Winning Price}
In this section we study the association between set-asides and winning prices among large and small bidders using the same weighted least squares model as in equation 6. The only differences are that we use the log of winning price as the response variable and restrict the data to winning bids. 

Table 4 reports the results with heteroskedasticity-consistent robust standard errors. Again, OLS coefficients are close to WLS coefficients which we'll discuss in detail. Taking the exponential of the coefficients gives the corresponding percentage change in winning prices. Compared to Table 3, the coefficient of small bidder indicator is now insignificant, suggesting that for items without set-asides, small business winners and large business winners win at the same price, which is an outcome of standard first price auction. For items with a 50\% set-aside, large winners and small winners are still displaying price decreases, with small bidders dropping more. 50\% set-aside predicts a 7\% and 12.5\% decrease on average in large and small bidders' winning prices respectively. These coefficients have a smaller magnitude than those in the offer price model, potentially because there's less room for price adjustment on low prices and net profit drops faster at low prices. For items with a 100\% set-aside, small bidders' winning prices are expected to increase by 2.1\%. Total demand for a product is associated with a roughly linear increase in winning prices, ranging from 0.09\% to 16.8\%. Number of bidders predicts a decrease in winning prices as expected, with a similar nonlinear trend as in the offer price model. The model predicts that adding one more bidder for a typical item is expected to bring down the winning price by about 3\%. All the coefficients discussed above are significant unless otherwise noted.

\begin{table}
    \centering
    \begin{tabular}{r|cc|cc}
    \toprule
    & \multicolumn{2}{c|}{WLS} & \multicolumn{2}{c}{OLS} \\
    & Coefficient & Robust Error & Coefficient & Robust Error \\\midrule
    Constant & -1.0725 & (0.0814) & -1.1516 & (0.0803) \\
    Small & 0.0204 & (0.0210) & 0.0064 & (0.0183) \\
    SA50\%, Large & -0.0723 & (0.0284) & -0.0579 & (0.0243) \\
    SA50\%, Small & -0.1338 & (0.0234) & -0.1022 & (0.0192) \\
    SA100\%, Small & 0.0212 & (0.0090) & 0.0247 & (0.0090) \\
    Demand & 0.0243 & (0.0013) & 0.0233 & (0.0013) \\
    Demand$^2$ & -0.0009 & (0.0001) & -0.0008 & (0.0001) \\
    Number of bidders & -0.0324 & (0.0029) & -0.0305 & (0.0029) \\
    Number of bidders$^2$ & 0.0012 & (0.0002) & 0.0011 & (0.0002) \\
    log(USDA price) & 1.3396 & (0.0816) & 1.4042 & (0.0820) \\
    log(wholesale price) & 0.7956 & (0.0121) & 0.7949 & (0.0120) \\
    SDVOSB & 0.0290 & (0.0023) & 0.0301 & (0.0022) \\ \midrule
    $n=12677$ &\multicolumn{2}{c|}{$R^2=0.4501$} & \multicolumn{2}{c}{$R^2=0.4471$} \\ \bottomrule
    \end{tabular}
    \caption{Log of Winning Price}
\end{table}

\section{Discussion}
We have shown that 50\% set-asides in USDA procurement auctions are associated with a decrease in prices among large and small bidders. For small bidders in particular, the price drop can be as large as 21.7\% and 12.5\% in their offer prices and winning prices respectively, and it immediately follows that their price margins will decrease by a larger percentage. Recall from Figure 2 right panel that plots the share of quantities won by small bidders in total, there's little room for improvement regarding small businesses' overall probability of winning. Putting these evidences together, it's very unlikely that 50\% set-asides would result in an increase in small bidders' net profit in general, but it's still possible that some small businesses could benefit at the cost of other small and large businesses.

A potential issue with our regressions is that the numbers of observations associated with different product types are unbalanced, which may lead to over-representation of certain products and under-representation of the others. To resolve this concern, Appendix A replicates Table 2, 3 and 4 using a different weight in the least squares model. The new weight is such that within each product type, item weights are proportional to their quantity. We then scale the weights in each product type by the total quantity of that product, resulting in equal total weights for all product types. Standard errors are larger but significance level do not change. Regression coefficients change moderately in scale, but their sign and relative magnitude are consistent with Table 2, 3 and 4, suggesting that we have effectively controlled for differences among products and our predictions are not biased by product heterogeneity.

Looking forward, there are several questions that can be explored in future work. The first question is whether some small businesses, potentially the cheaper and more developed ones, are benefiting from USDA's set-aside rules at the cost of sacrificing higher-cost small businesses. Set-asides focus on selecting small businesses in the lower price quantile to meet a quota, which might not benefit the most disadvantaged small businesses in practice. Flat rate or percentage subsidy is a possible alternative to consider since the extent to which small bidders benefit upon winning is relatively fixed. Another question is whether bidders differ in bidding strategy not only because of business type but also because some bidders are more experienced and have more resources dedicated to implementing the optimal strategy. Strategizing in such a complicated auction could be challenging to newcomers and small firms. A question immediately follows that our models treat bidders at all time equally, though it's possible that they learn and adapt their strategies through these highly repetitive sequential auctions. We hope to engage in some of these questions in future research.

\section{Conclusion}
This paper studies the implications of small business set-asides policy in USDA food procurement auctions, discerning partial set-asides and full set-asides. Using five years of bid prices data on different beef products procured by the USDA, we've shown that 50\% set-aside is associated with a decrease in offer prices and winning prices among both large and small bidders. In a setting like the USDA where entry of new bidders is difficult, applying set-asides has little power in attracting more small bidders, yet large bidders are pushed by unfavorable policy to bid aggressively, and then small bidders are also affected. In addition, partial set-asides increase the uncertainty faced by all bidders because allocation of an individual item will depend on all items of the same product type. In 100\% set-asides, we do observe a price increase among small bidders likely because competition is relieved. But the magnitude is small because there are few potential large competitors anyway. The impact of set-asides policies rely on various aspects of the setting of the auction, and agencies need case-by-case investigation to work out the optimal policy. 

\newpage
\bibliographystyle{apacite}
\bibliography{Bibliography}

\newpage
\begin{appendices}
\section{Replicate Table 2, 3 and 4 equally weighting product types}
\begin{table}[H]
    \centering
    \begin{tabular}{r|cc}
    \toprule
    & \multicolumn{2}{c}{WLS} \\
    & Coefficient & Robust Error \\\midrule
    Constant & 0.3995 & (0.0322) \\
    Small & 1.5193 & (0.0288) \\
    SA50\%, Large & -0.7710 & (0.0919) \\
    SA50\%, Small & 1.8050 & (0.0904) \\
    SA100\%, Small & 0.1864 & (0.0381) \\
    Demand & 0.2259 & (0.0163) \\
    Demand$^2$ & -0.0083 & (0.0011) \\
    SDVOSB & 0.4870 & (0.0584) \\ \midrule
    $n=26169$ &\multicolumn{2}{c}{$R^2=0.8197$} \\ \bottomrule
    \end{tabular}
    \caption{Replicating Number of Bidders Model (Table 2)}
\end{table}
\begin{table}[H]
    \centering
    \begin{tabular}{r|cc}
    \toprule
    & \multicolumn{2}{c}{WLS} \\
    & Coefficient & Robust Error \\\midrule
    Constant & -0.8150 & (0.0388) \\
    Small & 0.0212 & (0.0109) \\
    SA50\%, Large & -0.1802 & (0.0142) \\
    SA50\%, Small & -0.2392 & (0.0089) \\
    SA100\%, Small & 0.0124 & (0.0067) \\
    Demand & 0.0148 & (0.0007) \\
    Demand$^2$ & -0.0005 & (0.0000) \\
    Number of bidders & -0.0324 & (0.0021) \\
    Number of bidders$^2$ & 0.0013 & (0.0001) \\
    log(USDA price) & 1.3773 & (0.0344) \\
    log(wholesale price) & 0.7357 & (0.0090) \\
    SDVOSB & 0.0211 & (0.0018) \\ \midrule
    $n=96474$ &\multicolumn{2}{c}{$R^2=0.4633$} \\ \bottomrule
    \end{tabular}
    \caption{Replicating Log of Offer Price Model (Table 3)}
\end{table}
\begin{table}[H]
    \centering
    \begin{tabular}{r|cc}
    \toprule
    & \multicolumn{2}{c}{WLS} \\
    & Coefficient & Robust Error \\\midrule
    Constant & -1.2423 & (0.2137) \\
    Small & -0.0099 & (0.0185) \\
    SA50\%, Large & -0.0527 & (0.0279) \\
    SA50\%, Small & -0.1058 & (0.0253) \\
    SA100\%, Small & 0.0197 & (0.0113) \\
    Demand & 0.0246 & (0.0039) \\
    Demand$^2$ & -0.0011 & (0.0002) \\
    Number of bidders & -0.0245 & (0.0069) \\
    Number of bidders$^2$ & 0.0009 & (0.0005) \\
    log(USDA price) & 1.7384 & (0.2090) \\
    log(wholesale price) & 0.6616 & (0.0390) \\
    SDVOSB & 0.0157 & (0.0039) \\ \midrule
    $n=12677$ &\multicolumn{2}{c}{$R^2=0.6129$} \\ \bottomrule
    \end{tabular}
    \caption{Replicating Log of Winning Price Model (Table 4)}
\end{table}
\end{appendices}

\end{document}